\documentclass[twocolumn,prb,superscriptaddress]{revtex4}

\usepackage[dvips]{graphicx}
\usepackage{amsmath}
\usepackage{pslatex}

\newcommand{\ehat}{\widehat{\mathbf{E}}_M}
\newcommand{\qwgm}{Q_M}
\newcommand{\gammae}{\gamma_e(0)}

\begin{document}

\pagenumbering{arabic}

\title{Actuation of Micro-Optomechanical Systems Via Cavity-Enhanced Optical Dipole Forces}

\author{Matt Eichenfield}
\email{matt@caltech.edu}
\author{Christopher P. Michael}
\author{Raviv Perahia}
\author{Oskar Painter}
\affiliation{Thomas J. Watson, Sr. Laboratory of Applied Physics, California
Institute of Technology, Pasadena, CA}

\date{\today}

\begin{abstract}
We demonstrate an optomechanical system
  employing a movable, micron-scale waveguide optically-coupled to a high-$Q$ optical microresonator.  We show that
  milliwatt-level optical powers create micron-scale displacements of the input waveguide. The displacement is caused by
  a cavity-enhanced optical dipole force (CEODF) on the waveguide, arising from the stored optical field of the
  resonator.  The CEODF is used to demonstrate tunable cavity-waveguide coupling at sub-mW input powers, a form of
  all-optical tunable filter.  The scaling properties of the CEODF are shown to be independent of the intrinsic
  $Q$ of the optical resonator and to scale inversely with the cavity mode
  volume.
\end{abstract}

\maketitle

\noindent Although light is usually thought of as
  imponderable, carrying energy but relatively little momentum, light can exert a large force per photon if confined to
  small structures.  Such forces have recently been proposed~\cite{ref:Povinelli1,ref:Povinelli2} as a means to
  construct novel optomechanical components such as tunable filters, couplers, and lasers.  Other theoretical
  studies of the nonlinear dynamics of these systems have shown them to be useful for performing optical wavelength
  conversion and efficient optical-to-mechanical energy conversion\cite{ref:Notomi5,ref:Notomi4}.  In the field of
  quantum physics, there has also been recent interest in using radiation pressure forces within micro-optomechanical
  resonators to help cool macroscopic mechanical oscillators to their quantum-mechanical ground
  state~\cite{ref:Gigan1,ref:Kleckner,ref:Arcizet,ref:Schliesser}.  Here, we demonstrate an optomechanical system
  employing a movable, micron-scale waveguide optically-coupled to a high-$Q$ optical microresonator.  We show that
  milliwatt-level optical powers create micron-scale displacements of the input waveguide. The displacement is caused by
  a cavity-enhanced optical dipole force (CEODF) on the waveguide, arising from the stored optical field of the
  resonator.  The CEODF is used to demonstrate tunable cavity-waveguide coupling at sub-mW input powers, a form of
  all-optical tunable filter.  Finally, the scaling properties of the CEODF are shown to be independent of the intrinsic
  $Q$ of the optical resonator, and to scale inversely with the cavity mode volume, indicating that such forces may
  become even more effective as devices approach the nanoscale.

\begin{figure}[t]
\begin{center}
\includegraphics[width=\columnwidth]{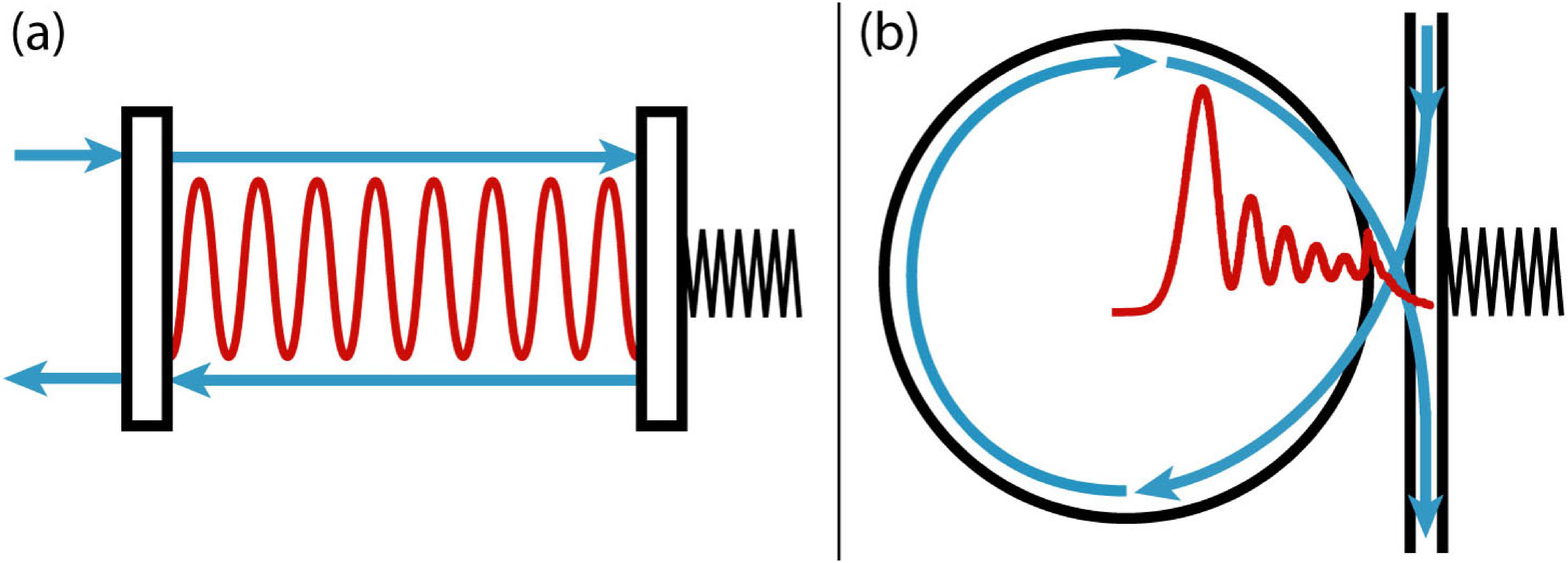}
\caption{\textbf{Schematic diagrams of two different optomechanical cavity systems}. Blue arrows indicate the propagation of light, with the resulting intensity
  profiles in the resonators shown as red lines.  \textbf{a}, Fabry-Perot cavity system with the position of the back
  mirror behaving as a mechanical oscillator. Radiation pressure from the stored cavity field pushes on the reflective
  back mirror, causing a power-dependent nonlinearity in the length of the cavity.  \textbf{b}, Evanescently-coupled
  cavity system.  The position of the input waveguide acts as a mechanical oscillator in this case.  A CEODF from the
  stored internal field of the resonator attracts the input waveguide, causing a power-dependent nonlinearity in the
  coupling of the waveguide to the cavity.} \label{fig:schem_comp}
\end{center}
\end{figure}

The ponderomotive effects of light within optical resonators have long been considered in the field of high-precision
measurement\cite{ref:Braginsky3}.  The canonical system, shown in
Fig.~\ref{fig:schem_comp}a, consists of a Fabry-Perot (FP) resonant cavity formed between a rigid mirror and a movable
mirror attached to a spring or hung as a pendulum\cite{ref:Dorsel}. A
nearly-resonant optical field builds up in amplitude as it bounces back-and-forth between the mirrors and pushes on the
movable mirror with each reflection, which detunes the FP cavity. The nonlinear dynamics associated with the
displacement of the mirror and the build-up of internal cavity energy result in an ``optical spring''
effect\cite{ref:Meystre}.  Under conditions in which the optical field cannot adiabatically follow the mirror movement, the radiation pressure force can drive or dampen oscillations of the position of the movable mirror\cite{ref:Pai,ref:Kippenberg4,ref:Marquardt1}---this effect is the basis for some optomechanical cooling schemes\cite{ref:Gigan1,ref:Arcizet,ref:Schliesser}.

In contrast to the FP optomechanical system, the system studied here consists of a monolithic, whispering-gallery-mode
(WGM) resonator coupled to an external waveguide.  The waveguide is suspended (secured at two distant
points) and behaves as if attached to a spring. Light evanescently couples into the resonator from the waveguide, as illustrated in Fig.~\ref{fig:schem_comp}b.  The intra-cavity light intensity changes the waveguide position via an all-optical
force on the waveguide due to the field of the resonator.  The resulting movement changes the waveguide-resonator
coupling-rate---rather than the cavity resonance condition as in the FP system---which is sensitive to the distance
between the waveguide and resonator. Unlike the FP system, the optical force is derived from the gradient force, a
result of intensity-dependent light shifts of electronic states in the dielectric external
waveguide\cite{ref:Gordon1,ref:Metcalf_van_der_Straten}.  No complete correspondence between
the FP system and the system in Fig.~\ref{fig:schem_comp}b can be made due to the external nature of the waveguide (the
FP would require a mirror that could change its reflectivity in response to changing optical power\cite{ref:Fermann}).
This non-trivial difference also applies to a wide class of cavity geometries in which a non-resonant dielectric object loads the cavity.

\begin{figure}[t]
\begin{center}
\includegraphics[width=\columnwidth]{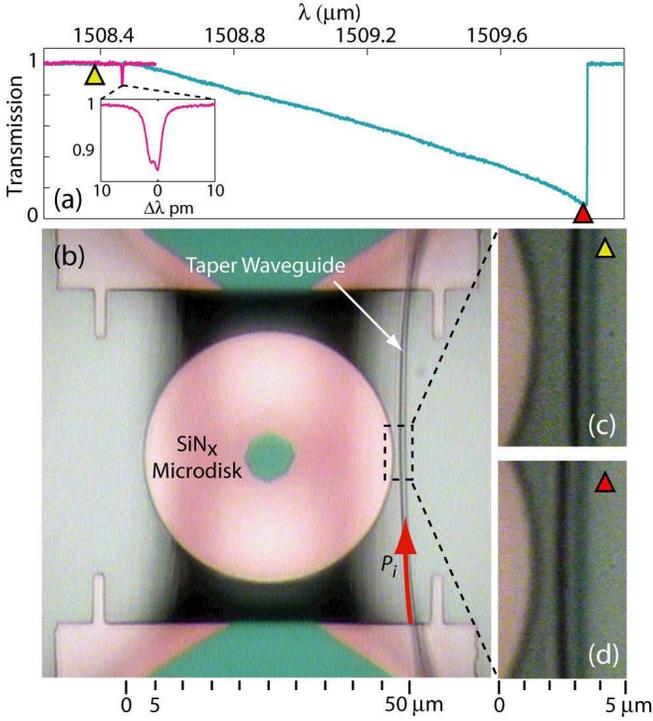}
\caption{\textbf{Waveguide displacement.} \textbf{a}, (inset) Low
power ($< 1$ $\mu$W) linear transmission spectrum of a high-$Q$ WGM
of the microdisk (the slight ``doublet'' character of the mode spectrum is a result of surface-scattering-induced modal coupling of degenerate clockwise and counterclockwise WGMs\cite{ref:Borselli2}). High power ($2.5$ mW) transmission spectrum (same
mode as inset) showing thermo-optic
  bistability and a change in on-resonance transmission depth.  The yellow(red) marker indicates the
  transmission level in panel c(d).  \textbf{b}, Low-magnification, top-view image showing the microdisk-taper
  system.  \textbf{c}, High-magnification image of the taper-disk coupling region taken at low optical input
  power, while the system is undercoupled.  \textbf{d}, High-magnification image taken at high input
power near resonance and critical coupling, showing a significant
deflection
  of the taper waveguide towards the disk resonator.}
\label{fig:displacement}
\end{center}
\end{figure}

In this work, the implementation of the optomechanical system of Fig.~\ref{fig:schem_comp}b consists of a high-$Q$
silicon nitride (SiN$_{x}$) microdisk resonator of diameter $D=44.8$ $\mu$m and thickness $t=253$ nm.  The resonator is
fed optical power through an external waveguide formed from a micron-scale silica optical fiber taper.  An optical
micrograph of the micron-scale fiber taper in the near-field of the SiN$_{x}$ microdisk is shown in
Fig.~\ref{fig:displacement}b, indicating the geometry of the taper-disk coupling.  Fabrication of the microdisk and the
taper are described in detail elsewhere\cite{ref:Barclay8,ref:Michael}.

Figure~\ref{fig:displacement}a shows the transmission spectrum of a high-$Q$ WGM of the microdisk at low (inset) and
high power. At low powers ($< 1$ $\mu$W), the microdisk system behaves linearly. At higher powers, heating effects due
to linear optical absorption within the SiN$_{x}$ microdisk result in thermo-optic bistability. In addition to the
thermo-optic effect (which gives rise to an asymmetric, shark-fin-like feature in the transmission
spectrum~\cite{ref:Carmon}), the on-resonance transmission depth changes significantly with input power. This effect can
be visually correlated with movement of the waveguide on a magnified microscope image such as
Fig.~\ref{fig:displacement}c-d.  Displacements on the order of the diameter of the waveguide ($1.1$~$\mu$m) can be
produced with an incident power of $2.5$~mW.  The waveguide always moves towards the edge of the microdisk, with the
movement disappearing when the waveguide is positioned far from the microdisk resonator.

We model the force by considering the potential energy, $\Phi$, of the waveguide due to the polarization induced by the
field of the resonator. Using complex fields and assuming linear dielectric susceptibility, the time-averaged
polarization energy of the waveguide due to a dipole moment per unit volume $\mathbf{P}(\mathbf{r})$ is

\begin{equation}\label{equ:pol_energy_general}
\Phi = -\frac{1}{2}\int\limits_{V_{\mathrm{wg}}}
\mathbf{P}(\mathbf{r})\cdot\mathbf{E}_M^\ast(\mathbf{r})\;\mathrm{d}^3\mathbf{r}
= -\frac{\epsilon_0 \chi_{WG}}{2}\int\limits_{V_{\mathrm{wg}}}
\left|\mathbf{E}_M(\mathbf{r})\right|^2\,\mathrm{d}^3\mathbf{r} \; ,
\end{equation}

\noindent where $V_{\mathrm{WG}}$ is the geometric volume of the
dielectric waveguide, $\chi_{WG}$ is the waveguide material's electric susceptibility and $\mathbf{E}_M(\mathbf{r})$ is
the electric field of the microdisk WGM. The average stored cavity energy, $U$, can be related to an effective mode volume of the
WGM, $V'_M = 2U/n_M^2\max\left[\left|\mathbf{E}_M(\mathbf{r})\right|^2\right]$,
%
%
where $n_M$ is the refractive index of the resonator. All
the information of field amplitudes is carried in $U$ and $V'_M$,
which allows one to write

\begin{equation}\label{equ:pol_energy_unevaluated}
\Phi = -\frac{\chi_{WG}\,U}{n_M^2 V'_M}\int\limits_{V_{\mathrm{WG}}}
\left|\ehat(\mathbf{r})\right|^2\mathrm{d}^3\mathbf{r}\;,
\end{equation}

\noindent where $\ehat(\mathbf{r})$ is the unitless electric field
eigenfunction of the WGM, normalized such that
$\max\left[\left|\ehat(\mathbf{r})\right|\right]=1$.  From FEM
simulations, we find $V'_M = 159 (\lambda_0/n_{M})^3$ for the WGM
 of resonance wavelength $\lambda_{0} = 1508.5$ nm studied here (see Methods).

In the limit of small parasitic losses\cite{ref:Barclay7}, the steady-state internal energy in the microdisk resonator is given by, $U = (1- T)\qwgm P_{i}/\omega$,
%
%
where $P_{i}$ is the input power in the waveguide at the resonator; $\omega$ is the angular frequency of the input
light; $\qwgm$ is the intrinsic quality factor of the WGM; and $T$ is the fractional waveguide transmission past the
cavity.  Fitting the spectrum of the WGM yields an unloaded, intrinsic $Q$-factor of $\qwgm=1.1\times 10^6$.

\begin{figure}[t]
\begin{center}
\includegraphics[width=\columnwidth]{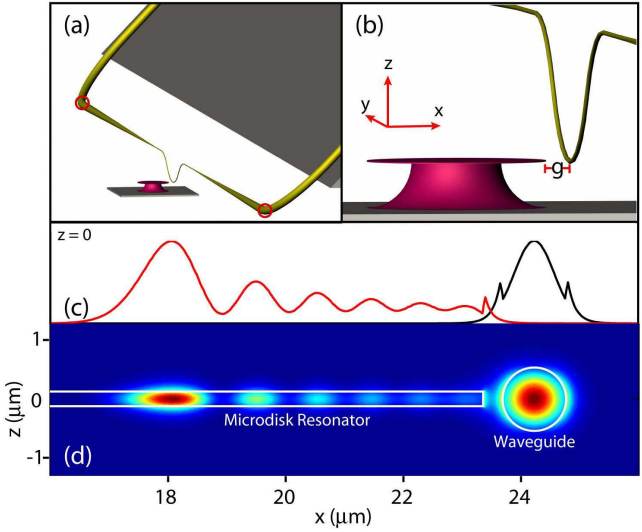}
\caption{\textbf{Physical structure and fields of the
waveguide-resonator system.} \textbf{a}, Three-dimensional schematic
view of the waveguide-resonator system. The tapered waveguide
behaves roughly like a guitar string, pinned at the corners of the
``U'' (marked by red circles). \textbf{b}, Zoomed-in view of the
waveguide-resonator system with coordinate reference. \textbf{c},
Projection of $|\ehat(\mathbf{r})|^2$ along the $x$-axis for the WGM
(red) and waveguide mode (black). \textbf{d}, Color-contour plot of
$|\ehat(\mathbf{r})|^2$ in the $x$-$z$ plane; the physical
boundaries are highlighted in white.}
\label{fig:modes_and_fields}
\end{center}
\end{figure}

\begin{figure}[t]
\begin{center}
\includegraphics[width=\columnwidth]{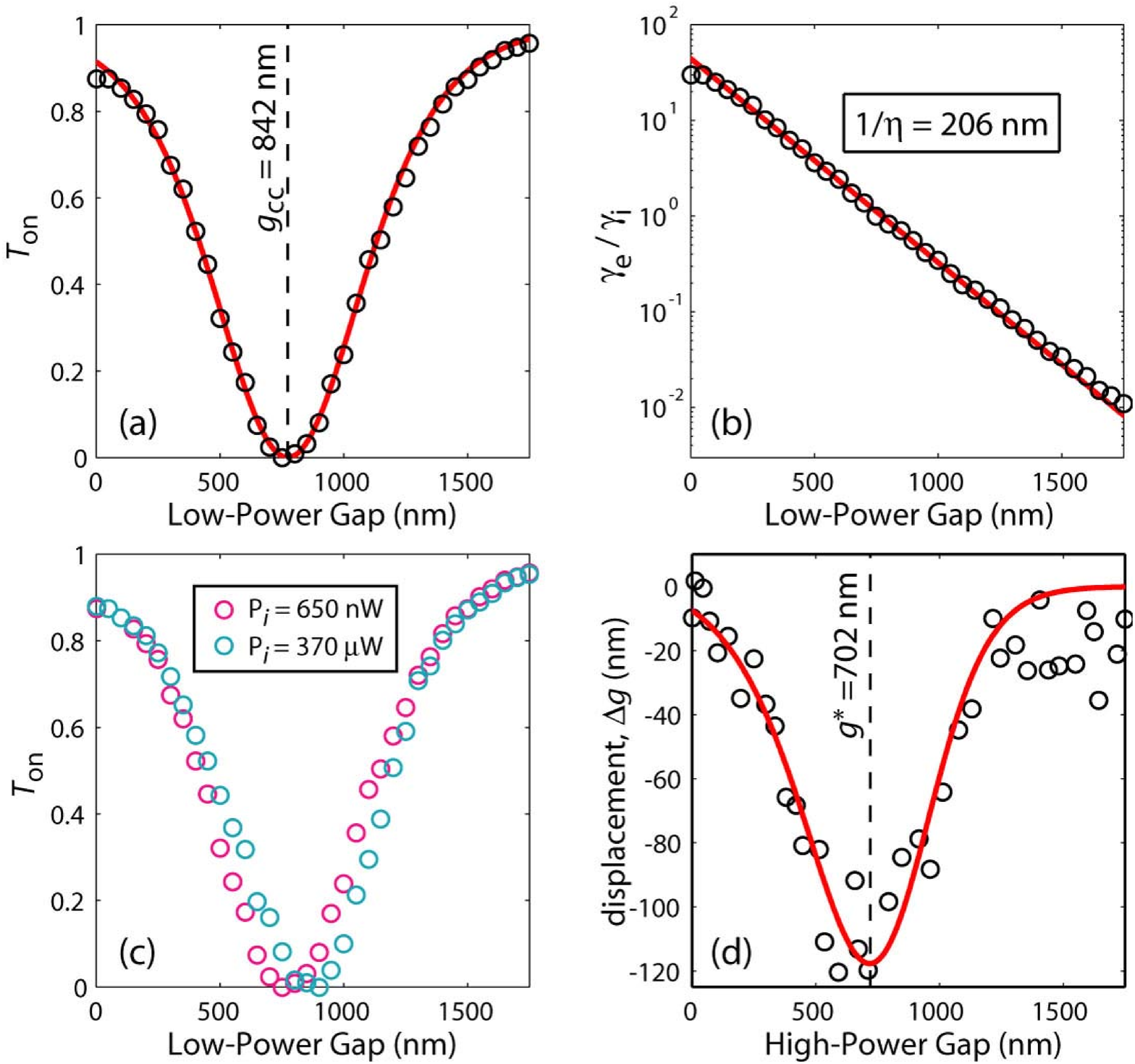}
\caption{\textbf{Characterization of optical coupling and
measurement of the CEODF.} \textbf{a}, Measured transmission curve at low power,
$T_{\text{on}}$ vs. $g$, for the microdisk-taper system
(a fit assuming exponential dependence of coupling on gap is shown as
a solid blue line). \textbf{b}, $\gamma_e/\gamma_i$ and fit
extracted from (a) showing exponential dependence of coupling on
gap. \textbf{c}, $T_{\text{on}}$ vs. low-power gap for high (cyan)
and low (magenta) input powers. \textbf{d}, Inferred $\Delta g$ vs.
high-power gap (equivalent to $F_{\text{o}}/k_{s}$). The actual
high-power taper-disk gap is extracted from the fit to the low power
coupling curve in panel a.  The fit theoretical model of the CEODF
is shown as a solid red line.} \label{fig:data}
\end{center}
\end{figure}

The on-resonance transmission ($T_{\text{on}} = T(\omega_0)$)
depends on the cavity-waveguide optical coupling- and loss-rates, $T_{\text{on}} = \left[(1-\gamma_e/\gamma_i)/(1+\gamma_e/\gamma_i)\right]^2$,
%
%
where $\gamma_e$ is the extrinsic (waveguide-to-resonator) coupling-rate, and $\gamma_i = \omega_0/\qwgm$ is the
intrinsic cavity loss-rate\cite{ref:Spillane2}.  The fields of both the microdisk resonator and the taper waveguide
decay exponentially outside of their geometric boundaries due to the evanescent nature of the fields there (see Figs.
\ref{fig:modes_and_fields}(c-d)).  We thus expect $\gamma_e$ to vary exponentially with the disk-taper gap, $g$ (shown
in Fig.~\ref{fig:modes_and_fields}b); i.e.  $\gamma_e(g)=\gamma_e(0)\exp(-\eta g)$, where $\gamma_e(0)$ is nominally the
``zero-gap'' coupling-rate.  In Fig.~\ref{fig:data}b we plot $\gamma_e(g)$ as extracted from the measured
$T_{\text{on}}(g)$ curve in Fig.~\ref{fig:data}a and unloaded $Q$-factor of the microdisk.  $\gamma_e(g)$ shows the
expected exponential dependence with gap, yielding a decay constant equal to $\eta = 1/206$ nm$^{-1}$ for the disk-taper
system.  The fully-characterized low-power transmission curve (Fig.~\ref{fig:data}a), can be used to infer the relative
disk-taper gap from any value of on-resonance transmission, $T_{\text{on}}$, given knowledge of whether the system is
overcoupled or undercoupled.

Substituting in the exponential dependence of coupling on gap, one
arrives at a dipole potential energy of the waveguide given by,

\begin{equation}\label{equ:pol_energy_unevaluated_T_substituted}
\Phi = -\frac{4 \chi_{\text{WG}} \, \gammae Q^2_M \,
P_{i}}{\omega^2_0 \, n_{M}^2 V'_M}\frac{e^{- \eta
g}}{\left(1+\frac{\gammae \qwgm}{ \omega_0} e^{-\eta g}\right)^2}
\int\limits_{V_{\text{WG}}}
\left|\ehat(\mathbf{r})\right|^2\mathrm{d}^3\mathbf{r}\;.
\end{equation}

\noindent The integral in equation (\ref{equ:pol_energy_unevaluated_T_substituted}) can be approximated by $V'_{\text{WG}}\, e^{-\sigma g}$,
%
%
where $\sigma$ is an effective decay constant, and $V'_{\text{WG}}$ is an effective volume of the waveguide. In what
follows, we take $\sigma \approx \eta$, as predicted by a simple coupled-mode theory of the taper-microdisk
system\cite{ref:Manolatou} (as shown below, this assumption is borne out by experiment).

Although the amount of externally-coupled energy changes with the
gap, the fast response rate of the optical cavity ($ 0.88$~ns photon
lifetime) allows the resonator dynamics to be adiabatically removed.
In this approximation, $\Phi(g)$ describes a completely conservative
energy landscape without hysteretic or dissipative characteristics.
The effective optical force on the waveguide is,

\begin{equation}\label{equ:force}
F_{\text{o}}(g) = -\frac{\partial\Phi(g)}{\partial g} = -\frac{8
\eta V'_{\text{WG}} \chi_{\text{WG}} \, \gammae \, Q^2_M \, P_{i}}{\omega^2_0 \, n_{M}^2 V'_M}\frac{e^{ \eta g}}{\left(e^{
\eta g}+\frac{\gammae Q_M }{\omega_0} \right)^3} \;.
\end{equation}

\noindent  We note that this force
curve is significantly different from that of a closed system in which the optical dipole force would be given by the
product of the stored cavity energy and the gradient of the microdisk mode's evanescent near-field intensity profile.
The maximum of this force occurs at a gap $g^*=\eta^{-1}\ln[\gammae Q_M/2\omega_0]$, and is given by

\begin{equation}\label{equ:fmax}
F^{*}_{\text{o}}=F_{\text{o}}(g^*)=-\frac{32 \eta \,
\chi_{\text{WG}}}{27 n^2_{M} \gammae} \frac{V'_{\text{WG}}}{V'_M}
P_{i}\;,
\end{equation}

\noindent In this model, $g^*$ occurs on the overcoupled side of the
critical coupling gap---not at $g_{\text{cc}}=\eta^{-1}\ln[\gammae
Q_M/\omega_0]$, the point of maximum stored cavity energy.

To test the validity of the model, we measure the position-dependence of the steady-state force.  Starting with the
waveguide far from the microdisk resonator ($g>1750$ nm), we incrementally move the resonator toward the waveguide in 50
nm steps using an encoded DC motor stage. At each position of the motor, we measure $T_{\text{on}}$ at 650 nW and 370
$\mu$W.  Figure~\ref{fig:data}c shows the high- (cyan) and low-power (magenta) transmission curves.  The actual position
of the waveguide in the high-power measurement is extracted from the low-power $T_{\text{on}}$ data.  The change in
actual waveguide-resonator gap, $\Delta g$, between the low- and high-power transmission curves is proportional to the
force required to move the waveguide to its high-power position, assuming the mechanical spring constant ($k_s$) of the
optomechanical system is linear. Thus, Fig.~\ref{fig:data}d is a plot of $F_{\text{o}}(g)/k_{s}$.

Figure~\ref{fig:data}d shows the fit to the force curve data using equation~(\ref{equ:force}).  As mentioned above,
coupled mode theory predicts that $\sigma \approx \eta$; this equality is confirmed here as $\sigma/\eta = 1.01 \pm
0.03$ is the best non-linear least-squares fit to the measured force curve when $\sigma$ is left as a free parameter.
The fit model also matches the measured recovery to zero force for small gaps, and has a peak force position ($g^* =
702$~nm), just on the overcoupled side of the zero transmission gap ($g_{\text{cc}} = 842$~nm).  The numerical
evaluation of the optical force in equation~(\ref{equ:force}) yields a prediction for the magnitude of the optical force
on the taper waveguide and, from the fit of the measured force curve in Fig.~\ref{fig:data}d, an estimate of the taper's
mechanical spring constant.  We find a peak optical force of 20 pN/mW of input power for the disk-taper system under
study and a taper spring constant of $k_{s} = 73$ pN/$\mu$m.

Direct measurements of the mechanical properties of the taper
waveguide were also performed.  A diagram illustrating the geometry
of the fiber taper waveguide used in these experiments is given in
Figs. \ref{fig:modes_and_fields}(a-b). Imaging of the fiber taper
displacement with a microscope indicates that the optically-excited
mechanical resonance of the taper is similar to the fundamental mode
of an elastic wire, pinned at the corners of the ``U"-shaped bend in
the taper waveguide.  As shown, the fiber taper waveguide is not a
simple ``U'' shape, but rather has a small ``dimple'' ($\sim 160$
$\mu$m radius of curvature) at the taper's center to allow local
probing of planar devices\cite{ref:Michael}. The torsional mode of
the ``dimple'' does not appear to be excited.  From the measured
mechanical frequency ($\Omega=2 \pi \times 193$~Rad/s) and the
physical mass of the taper's ``U'' section ($m = 2.6\times10^{-11}$
kg), we arrive at $k_{s} \approx 38$~pN/$\mu$m (see Methods). This
value differs by less than a factor of two from the
optically-measured spring constant of the taper waveguide.

\begin{figure}[t]
\begin{center}
\includegraphics[width=\columnwidth]{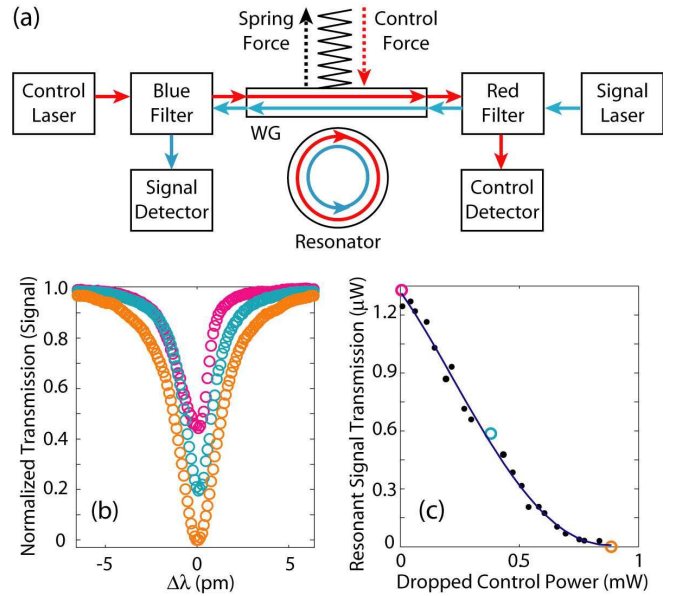}
\caption{\textbf{Optically tunable filter demonstration.} \textbf{a},
Experimental setup of the filter demonstration.
\textbf{b}, Transmission spectrum of the signal field for three
different dropped control powers. \textbf{c}, On-resonance
transmitted signal power vs. dropped control power (with trend
line).  The signal extinction ratio is $21$~dB for $~0.9$~mW of control
power. Transmission spectra in panel b are taken at the dropped
control powers indicated by open circles of matching color.}
\label{fig:modulation}
\end{center}
\end{figure}

Applications of the CEODF include optical control of micro- and nano-optomechanical devices for the dynamic switching,
filtering, or modulation of light.  Figure~\ref{fig:modulation}a shows the schematic of a simple experiment to
demonstrate all-optical tuning of the filter characteristics of the microdisk using the CEODF.  In this scheme a red
control field ($\lambda_{\mathrm{C}} \approx 1527 \textrm{nm}$) counterpropagates with a blue signal field
($\lambda_{\mathrm{S}} \approx 1492 \textrm{nm}$), each of the fields resonant with a different mode of the microdisk
(in this case, WGMs of the same radial order but azimuthal mode number difference $\delta m=4$).  The waveguide is
pulled towards the resonator in proportion to the dropped control power, tuning the cavity-waveguide system from the
undercoupled regime towards the overcoupled regime.  In this way, the filter rejection level at the signal wavelength
can be tuned by the control power level, as shown in Figs. \ref{fig:modulation}(b-c). Using the thermo-optic bistability
of the system, variation in the dropped control power is accomplished not by tuning the control power level, but rather
by tuning the \textit{wavelength} of the control laser.  This provides an approximately linearly-tunable control;
however, the resonance position for the signal is tuned simultaneously and must be tracked. By engineering a similar
system in which the threshold power for thermo-optic bistability is much larger than the threshold power for
optomechanical modulation, the same basic structures could be made to tune the cavity's filter properties by adjusting
only the input power of the control---without the need to track the signal resonance.  Utilizing a microfabricated
on-chip waveguide\cite{ref:Lee_MC}, an equivalent system with high mechanical frequency (MHz to
GHz\cite{ref:Huang_XMH,ref:Carmon3}) could be used to produce fast, all-optical tuning or signal modulation.

Analysis of equation~(\ref{equ:fmax}) indicates that the magnitude of the force does not scale with the intrinsic
optical $Q$-factor ($Q_M$). This is a result of the interplay between the stored optical energy and the degree of cavity
loading: increasing $Q_M$ moves $g^* \sim \ln[Q_M]$ further from the disk, which exactly cancels the $Q$-enhancement of
the internal field.  The $Q_M$-independence of the force facilitates making systems insensitive to other nonlinearities
which scale with $Q_M$ (such as the thermo-optic bistability of the current devices).  Indeed, early results with
low-$Q$ microdisks ($Q_M \sim 4100$) of roughly the same size ($D = 20$~$\mu$m) show large optical force effects with no
observable thermo-optic bistability. Equation (\ref{equ:fmax}) also shows that the force scales inversely with $V'_M$.
Therefore smaller devices, even with lower quality factors, give rise to a larger maximum force. Up to a factor of order
unity, $V'_{\text{WG}}/\gammae$ cancels, keeping the above scaling valid as devices approach the nano-scale.  For example,
in the case of planar two-dimensional photonic crystal nanocavities\cite{ref:Painter1} with cubic-wavelength-scale mode
volumes, the optical forces can be expected to be two orders of magnitude larger than in the microdisk studied here,
approaching the nN level.  Such scaling behavior is applicable to a wide variety of geometries in which the CEODF acts
on non-resonant dielectric objects in the near-field of the optical resonator and indicates an interesting path towards
low-power, high-speed micro- and nano-scale actuators and transducers.

\section*{Methods}

\begin{footnotesize}

  The fiber taper is created by locally heating (with a H$_{2}$ torch) and drawing a standard SMF-28e silica fiber.  An
  adiabatic taper is formed with a long ($4$ mm), narrow diameter region ($d\ = 1.1 \pm 0.1 \mu \mathrm{m}$) in which
  the evanescent field of the guided mode penetrates significantly into the air cladding.  The fiber is then bent in a
  ``U''-shape \cite{ref:Barclay5}, with the tapered region at the end of the ``U'' and the two fiber ends separated by
  an adjustable gap to provide varying tension levels to the taper. In addition, a ``dimpled" region is created in the
  geometric center of the ``U'' region in order to access the disk, which resides in a trench.  The fiber tension is
  mechanically adjustable; tension is optimized to accentuate optically-induced fluctuations while maintaining the
  waveguide's stability (the stability of the taper waveguide is exemplified in Fig.~\ref{fig:data}a).

  Fiber-pigtailed, external cavity, tunable lasers operating over the $1400$ and $1500$~nm wavelength bands are used to
  couple optical power into the taper waveguide. The polarization of the guided mode in the taper is controlled via
  three adjustable fiber loops. Excitation of the whispering-gallery modes (WGMs) of the microdisk is performed by
  positioning the fiber taper in the equatorial plane of the microdisk and approaching the near-field of the microdisk
  from the side.  Positioning of the disk ($x$-$y$) and taper ($z$) is accomplished using DC motors with $50 \pm 0.5$ nm
  encoded resolution.

  The low-power $T_{\text{on}}$ data in Fig.~\ref{fig:data}c allows the construction of a transmission curve that is
  identical to the transmission curve in Fig.~\ref{fig:data}a; it is nevertheless taken again, concurrently with the
  high power data, to cancel any drift in the taper-disk gap over the course of the measurement.

The particular WGM used in this study has six radial antinodes ($p=6$, $m=109$, $v=0$; shown in
Fig.~\ref{fig:modes_and_fields}c-d juxtaposed to the waveguide mode) and was chosen due to its high intrinsic (unloaded)
$Q$ and favorable taper-coupling characteristics (phase-matching).  Individual WGM modes can be
identified~\cite{ref:Borselli} via the following characteristics: the free-spectral range between modes, the
radiation-limited $Q$ of modes, and the strength of coupling to the taper waveguide (degree of phase-matching).  The
experimentally measured quantities are matched to accurate FEM simulations of the disk modes.  The mode's identification
was experimentally verified by moving the waveguide over the surface of the disk and measuring the characteristic radial
pattern of the mode through changes in the taper-disk coupling-rate, $\gamma_{e}$.

We measured the mechanical resonance frequency by driving the taper waveguide with an electro-acoustical transducer in
close proximity to the waveguide.  The response of the waveguide appears as a clearly-visible oscillation in the
on-resonance optical transmission of the cavity-waveguide system, well above the noise floor.  By measuring this
response against the frequency of the driving waveform, we extract the mechanical resonance frequencies.  The effective
mass of the oscillator is approximated by the mass of the fused-silica taper (length $8$ mm, diameter exponentially
tapering from $1.1$ to $10$ $\mu$m) weighted by the fundamental mechanical eigenfunction.

\end{footnotesize}

\section*{Acknowledgements}

\begin{footnotesize}

The authors sincerely thank Thomas Johnson, Paul Barlcay, and Kartik
Srinivasan for many fruitful discussions and helpful feedback.

\end{footnotesize}

%
%
%


\end{document}